\documentclass[aps,twocolumn,nofootinbib,superscriptaddress,floatfix,linenumbers]{revtex4}

\usepackage{color,amsmath,amssymb,graphicx,latexsym,subfigure}
\usepackage{threeparttable}
\usepackage[colorlinks]{hyperref}
\usepackage{lineno}

\newcommand{\beq}{\begin{equation}}
\newcommand{\eeq}{\end{equation}}
\newcommand{\bea}{\begin{eqnarray}}
\newcommand{\eea}{\end{eqnarray}}

\newcommand{\gsim}{\lower.7ex\hbox{$\;\stackrel{\textstyle>}{\sim}\;$}}
\newcommand{\lsim}{\lower.7ex\hbox{$\;\stackrel{\textstyle<}{\sim}\;$}}

\newcommand{\be}{\begin{equation}}
\newcommand{\ee}{\end{equation}}
\newcommand{\ba}{\begin{eqnarray}}
\newcommand{\ea}{\end{eqnarray}}

\begin{document}

\switchlinenumbers

\title{Is the $W$-boson mass enhanced by the axion-like particle, dark photon, or chameleon dark energy? }

\author{Guan-Wen Yuan}
\affiliation{Key Laboratory of Dark Matter and Space Astronomy, Purple Mountain Observatory, Chinese Academy of Sciences, Nanjing 210023, China}
\affiliation{School of Astronomy and Space Science, University of Science and Technology of China, Hefei 230026, China}

\author{Lei Zu\footnote{zulei@pmo.ac.cn}}
\affiliation{Key Laboratory of Dark Matter and Space Astronomy, Purple Mountain Observatory, Chinese Academy of Sciences, Nanjing 210023, China}

\author{Lei Feng\footnote{fenglei@pmo.ac.cn}}
\affiliation{Key Laboratory of Dark Matter and Space Astronomy, Purple Mountain Observatory, Chinese Academy of Sciences, Nanjing 210023, China}
\affiliation{School of Astronomy and Space Science, University of Science and Technology of China, Hefei 230026, China}

\author{Yi-Fu Cai}
\affiliation{School of Astronomy and Space Science, University of Science and Technology of China, Hefei 230026, China}
\affiliation{Department of Astronomy, School of Physical Sciences, University of Science and Technology of China, Hefei, Anhui 230026, China}
\affiliation{CAS Key Laboratory for Researches in Galaxies and Cosmology, University of Science and Technology of China, Hefei, Anhui 230026, China}

\author{Yi-Zhong Fan\footnote{yzfan@pmo.ac.cn}}
\affiliation{Key Laboratory of Dark Matter and Space Astronomy, Purple Mountain Observatory, Chinese Academy of Sciences, Nanjing 210023, China}
\affiliation{School of Astronomy and Space Science, University of Science and Technology of China, Hefei 230026, China}

\begin{abstract}

The $W$-boson mass ($m_{W}=80.4335 \pm 0.0094 \mathrm{GeV}$) measured by the Collider Detector at Fermilab collaboration is greater than the standard model (SM) prediction at a confidence level of $7\sigma$, strongly suggesting the presence of new particles or fields. In the literature, various new particles and/or fields have been introduced to explain the astrophysical and experimental data, and their presence, in principle, may also enhance the $W$-boson mass. 
In this study, we investigate axion-like particle (ALP), dark photon (DP), and chameleon dark energy (DE) models for a solution to the $W$-boson mass excess.
We find that the ALP and DP interpretations have been significantly narrowed down by global electroweak fits. The possibility of attributing the $W-$boson mass anomaly to the chameleon DE is ruled out by other experiments. 
\end{abstract}

\date{\today}


\maketitle
 
\section{Introduction}
With the discovery of the Higgs particles, the standard model (SM) \cite{Glashow:1961tr, Salam:1964ry, Weinberg:1967tq, Patt:2006fw} has achieved unprecedented success and interpreted almost all the phenomena measured by the colliders. However, astrophysical data, raises questions in the community. The observations of the galaxy rotation curves~\cite{Rubin:1980zd}, gravitational lensing~\cite{Clowe:2006eq}, and cosmic microwave background (CMB) power spectrum~\cite{Larson:2010gs} have demonstrated that the total mass of the universe cannot be accounted for by the ordinary matter in the SM, and dark matter (DM) has been introduced to explain the missing mass \cite{Bertone:2004pz, Feng:2010gw, 2018RvMP...90d5002B,Salucci:2018hqu}. The data of Ia supernova~\cite{SupernovaSearchTeam:1998fmf}, baryon acoustic oscillations~\cite{Beutler:2011hx}, large-scale structure~\cite{SDSS:2005xqv}, and CMB call for the presence of dark energy (DE) that makes up 68\% of the total energy density of the current universe \cite{Planck:2018vyg}. Though extensively explored, the nature of DM and DE is still essentially unknown. Various new particles or fields have been introduced in the literature~\cite{Cai:2009zp, Liu:2017drf, Jiao:2020gli, Zu:2020idx, Cao:2015efs, Yuan:2021mzi, Zhu:2022scj, Cao:2022mif}. Some of these particles or fields are also well theoretically motivated. For instance, axion, a hypothetical sub-eV particle beyond SM and a promising candidates for cold DM, has been initially introduced to solve the CP violation in strong interaction \cite{1977PhRvL..38.1440P, 1978PhRvL..40..223W}.

Notably, some anomalies have been reported in some physical experiments over the past few years. For example, in 2021, the muon anomalous magnetic moment was measured by E989 at Fermilab with a relative precision of 368 parts-per-billion. The combination with a previous measurement from Brookhaven National Laboratory yielded a deviation from the SM prediction at a confidence level (CL) of $4.2\sigma$ \cite{Muong-2:2021ojo}. The electroweak fit, including $Z$-pole data and $m_{\rm top}$ and $m_h$ measurements, yields a well-determined mass $m_{W, {\rm SM}}=80.361\pm 0.006$~GeV of the $W$-boson that mediates the electroweak interactions. Hence, the precision measurement of the $W$-boson mass is essential in probing the SM's internal consistency. Recently, the Collider Detector at Fermilab (CDF) Collaboration reported a precise measurement of $W$-boson mass \cite{CDF:2022hxs}, which is $m_{W}=80.4335 \pm 0.0094 \mathrm{GeV}$. The deviation from the SM prediction is approximately $\sim 7\sigma$~\cite{CDF:2022hxs}. 
This remarkable discovery is the focus of this study. The $W$-boson mass equals to $gv/2$ at the tree level, where $v {\rm = 245~GeV}$ is the vacuum expectation value of the Higgs field and $g$ denotes the weak-isospin coupling parameter. Like other particles in the SM, its mass also suffers from loop corrections. New particles/fields interacting with $W$-boson also have loop corrections to its mass. This fact motivates us to examine whether some DM/DE models could explain the latest CDF $W$-boson measurement. Accordingly, in this study, we investigate a few models, including the models of the axion-like particle (ALP), dark photon (DP) and the chameleon DE. 
We find out that the first two scenarios are still viable, whereas the last one is not. All models can enhance the loop corrections of the $W$-boson mass, but some parameters have already been ruled out by other experimental/astrophysical data. Although, the ALP and DP  (with $m_{Z ^\prime} \gg O(EW)$) models are still viable, we have identified some favored regions without violating other constraints.

\section{Global Electroweak Precision Fits} \label{stu_parameters}
The $W$-boson mass can be measured at a hadron collider through the charged current Drell-Yan process, i.e, $p\bar{p} \rightarrow W^{\pm}+X \rightarrow l^{\pm}+\nu(\bar{\nu})+X$. There are three observables in this process: the lepton-pair transverse mass, missing transverse momentum, and charged lepton transverse momentum. The measurement accuracy of the $W$-boson mass has improved rapidly with the growing experimental data and advancements in detection technology. Previously, the $W$-boson mass was measured to a precision of 19 MeV at ATLAS~\cite{ATLAS:2015ihy}, 23 MeV at CMS~\cite{CMS:2018ktx}, 59 MeV at LHCb~\cite{LHCb:2015jyu},
33 MeV at the Large Electron-Positron (LEP) collider \cite{ALEPH:2013dgf}, and 16 MeV at the Tevatron \cite{CDF:2013dpa} by averaging the measurements of CDF \cite{CDF:2012gpf} and D0 \cite{D0:2012kms}. The global fit result of the electroweak experiments available in 2018 is 7 MeV \cite{Haller:2018nnx}, which is 58\% of the 2021 PDG average of direct measurements \cite{ParticleDataGroup:2020ssz}.

The precision electroweak measurements in colliders involve fermions scattering each other. In these inelastic scattering processes, the new gauge bosons can be exchanged by fermions and induce an oblique correction, which would shift the SM coupling~\cite{Tsai:2015ugz, Stadnik:2015kia, Berlin:2018bsc}. When this correction dominates, we may use the freedom from new physics, particles/fields, to perform field redefinitions and put all new physics effects into the vacuum polarization ~\cite{Peskin:1991sw,Maksymyk:1993zm}.

For new physical models, the loop contribution to the $W$-boson mass can be described in terms of the oblique parameters of $\mathcal{(S,~T,~U)}$ \cite{Peskin:1991sw,Maksymyk:1993zm}, i.e.,
\begin{equation}
m_{W}^{2}=m_{W, \mathrm{SM}}^{2}+\frac{\alpha c_{W}^{2} m_{Z}^{2}}{c_{W}^{2}-s_{W}^{2}}\left[-\frac{\mathcal{S}}{2}+c_{W}^{2} \mathcal{T}+\frac{c_{W}^{2}-s_{W}^{2}}{4 s_{W}^{2}} \mathcal{U}\right],
\end{equation}
where $s_{\rm W}\equiv {\rm sin}\theta_{\rm W}, c_{\rm W}\equiv {\rm cos}\theta_{\rm W}$, and $\theta_{\rm W}$ denotes the Weinberg mixing angle. $\alpha$ and $m_Z$ are the fine structure constant and $Z$-boson mass, respectively. The SM-predicted $W$-boson mass is $m_{W, {\rm SM}} = 80.361\pm 0.006$GeV ~\cite{ParticleDataGroup:2020ssz}.
In addition, the complete electroweak precision measurements are parameterized by $\mathcal{S, T, U}$ that are given by \cite{Peskin:1991sw, Maksymyk:1993zm}. 

\begin{widetext}

\begin{table}[!t]
\setlength{\tabcolsep}{.5em}
\begin{tabular}{|c|c|ccc|c|ccc|c|ccc|}
\hline
  &  PDG2021~\cite{ParticleDataGroup:2020ssz} & \multicolumn{3}{c|}{Correlation } & Case A~\cite{deBlas:2022hdk} & \multicolumn{3}{c|}{Correlation}& Case B~\cite{Lu:2022bgw} & \multicolumn{3}{c|}{Correlation}\\
\hline
$S$ & $0.06 \pm 0.10$  & $1.00$ & &                  & $0.005 \pm 0.096$ & 1.00 & &                    & $0.06 \pm 0.10$ & $1.00$ & &\\
$T$ & $0.11 \pm 0.12$  & $0.90$ & $1.00$ &           & $0.04 \pm 0.12$ & $0.91$ & $1.00$ &             & $0.11 \pm 0.12$ & $0.90$ & $1.00$ & \\
$U$ & $-0.02 \pm 0.09$ & $-0.57$ & $-0.82$ & $1.00$  & $0.134 \pm 0.087$ & $-0.63$ & $-0.88$ & $1.00$  & $0.14 \pm 0.09$ & $-0.59$ & $-0.8$ & $1.00$\\
\hline
\end{tabular}
\caption{The values of the oblique parameters $S$, $T$, and $U$ and the correlation matrix allowed by global electroweak precision fit with the $W$-boson mass from PDF (2021)~\cite{ParticleDataGroup:2020ssz} and CDF (2022)~\cite{CDF:2022hxs}, respectively. Case A was adapted from Blas, et al~\cite{deBlas:2022hdk}, and Case B was adapted from Lu, et al~\cite{Lu:2022bgw}, the difference between Cases A and B is with and without considering the very recent measurement of the top quark mass by the CMS Collaboration~\cite{CMS-PAS-TOP-20-008}.
}
\label{STU_Fitted}
\end{table}

\end{widetext}

The global electroweak fit is a robust tool to explore the correlations among observables in the SM and predict the direction of new physics. Because the electroweak parameters in the SM are closely related to each other, we can expect that some observables in the global electroweak fits may suffer from new tensions once the $m_W$ is changed ~\cite{ALEPH:2005ab, Baak:2012kk, Baak:2014ora}. 
Two groups~\cite{Lu:2022bgw, deBlas:2022hdk}, replacing $80.379\pm 0.012$ GeV (PDF2021) with $80.4335\pm 0.0094$ GeV (CDF2022), have reported the oblique parameters, and Ref.~\cite{deBlas:2022hdk} used the very recent measurement of the top quark mass by the CMS Collaboration~\cite{CMS-PAS-TOP-20-008}, the results are presented in Table~\ref{STU_Fitted}.
We could find that these parameters are far more different from the old $\cal S$, $\cal T$, $\cal U$ values from PDG ~\cite{ParticleDataGroup:2020ssz} via a comparison between the new physics and experiments through the $\chi^2$ calculation based on $\mathcal{S, T, U}$ is~\cite{Baak:2014ora}:

\begin{widetext}
\begin{equation}
\chi_{}^{2}=\left(\begin{array}{lll}\Delta {\cal S} & \Delta {\cal T} & \Delta {\cal U} \end{array} \right) \left(\begin{array}{ccc}\sigma_{s}^{2} & C_{s t} \sigma_{s} \sigma_{t} & C_{s u} \sigma_{s} \sigma_{u} \\ 
C_{s t} \sigma_{s} \sigma_{t} & \sigma_{t}^{2} & C_{t u} \sigma_{t} \sigma_{u} \\ C_{s u} \sigma_{s} \sigma_{u} & C_{t u} \sigma_{t} \sigma_{u} & \sigma_{u}^{2}\end{array}\right) \quad\left(\begin{array}{c} \Delta {\cal S} \\ \Delta {\cal T} \\ \Delta {\cal U}\end{array}\right),
\label{chi2}
\end{equation}
\end{widetext}
where $(\sigma_s, ~\sigma_t, ~\sigma_u)$ are the errorbars,  $\Delta {\cal A}={\cal A}_{\rm model} - {\cal A}_0$, and ${\cal A}$ denotes $\mathcal{S, T, U}$. We could obtain the allowed parameter regions by $\Delta \chi^2 \equiv \chi^2 - \chi^2_{\rm min}$, and $\chi^2$ is dependent on the parameters in different new physical models. Below we discuss three specific models and set the constraints.

\section{New Physics and Constraints}\label{models}
\subsection{Axion-like Particle} 
ALP, which is a type of pseudoscalar particle, appear in string theory and some well-motivated models as a type of DM~\cite{Peccei:1977hh,Wilczek:1977pj,Preskill:1982cy,Abbott:1982af,Dine:1982ah,Hooper:2007bq,Marsh:2015xka,CAST:2017uph, Yuan:2020xui}. The difference between ALP and axion models is that both mass and coupling are free parameters for the former.
The effective interactions of ALP and electroweak gauge bosons, considered in this study, are given by the following dimension-5 effective Lagrangian~\cite{Bauer:2017ris, Yue:2021iiu, Bao:2022onq, Han:2022mzp}
\begin{equation}\label{axion_lagrangian}
\begin{aligned}
\mathcal{L}_{\mathrm{eff}}^{D \leq 5} =&\frac{1}{2}\left(\partial_{\mu} a\right)\left(\partial^{\mu} a\right)-\frac{m_{a}^{2}}{2} a^{2}+ g^2 C_{W W} \frac{a}{\Lambda} W_{\mu\nu}^{A} \tilde{W}^{\mu\nu, A}
\\ &+ g^{\prime 2} C_{B B} \frac{a}{\Lambda} B_{\mu\nu} \tilde{B}^{\mu\nu} ,
\end{aligned}
\end{equation}
where $W_{\mu\nu}^{A}$ and $B_{\mu\nu}$ denote the gauge field strengths of $SU(2)_L$ and $U(1)_Y$, respectively, and $g$ and $g^{\prime}$ denote the corresponding coupling constants, respectively.  $\tilde{B}^{\mu \nu}$ and $\tilde{W}^{\mu \nu}$ are their duals defined as $\tilde{B}^{\mu \nu}=\frac{1}{2} \epsilon^{\mu \nu \alpha \beta} B_{\alpha \beta}$ (with $\epsilon^{0123}=1$). The $\Lambda\sim 1$TeV is the energy scale of new physics. The ALP field and its mass are denoted by $a$ and $M_a$, respectively.
The interactions between ALP gauge bosons are contributed by the last two terms of Eq.~\eqref{axion_lagrangian}. 
In addition, the dimensionless couplings $C_{\gamma\gamma}= C_{WW} + C_{BB}$ describe the coupling strength of the interactions between ALP and photon in leading order. 

Because we only consider the ALP much lighter than the electroweak scale, where the effect from the ALP mass can be neglected, the corrections to electroweak precision observables can not be directly described by the  oblique parameters $\cal{S},\cal{T}$ and $\cal{U}$ in principle. In Ref.~\cite{Bauer:2017ris} the oblique parameters are equivalently given in terms of $\rho_*$, $s_W,s_0$ and $s_*$ by calculating the vacuum polarization function to one loop order (please see Subsection 6.3 therein for more details):

\begin{equation}
\begin{aligned} 
s_{*}^{2}=& \frac{g^{\prime 2}}{g^{2}+g^{\prime 2}}-s_{w} c_{w} \frac{\Pi_{\gamma Z}\left(m_{Z}^{2}\right)}{m_{Z}^{2}} \\ 
\rho_{*}=& 1+\frac{\Pi_{W W}(0)}{m_{W}^{2}}-\frac{\Pi_{Z Z}(0)}{m_{Z}^{2}}-\frac{2 s_{w}}{c_{w}} \frac{\Pi_{\gamma Z}(0)}{m_{Z}^{2}} \\ 
s_{W}^{2}=& \frac{g^{\prime 2}}{g^{2}+g^{\prime 2}}-c_{w}^{2}\left[\frac{\Pi_{W W}\left(m_{W}^{2}\right)}{m_{W}^{2}}-\frac{\Pi_{Z Z}\left(m_{Z}^{2}\right)}{m_{Z}^{2}}\right] \\ 
s_{0}^{2}=& \frac{g^{\prime 2}}{g^{2}+g^{\prime 2}}+\frac{s_{w}^{2} c_{w}^{2}}{c_{w}^{2}-s_{w}^{2}}\left[\frac{\Pi_{\gamma \gamma}\left(m_{Z}^{2}\right)}{m_{Z}^{2}}+\frac{\Pi_{W W}(0)}{m_{W}^{2}}\right]\\ 
 &-\frac{s_{w}^{2} c_{w}^{2}}{c_{w}^{2}-s_{w}^{2}}\frac{\Pi_{Z Z}\left(m_{Z}^{2}\right)}{m_{Z}^{2}}
\end{aligned}
\end{equation}

where the vacuum-polarization functions are defined by the decomposition $\Pi_{A B}^{\mu \nu}(q)=\Pi_{A B}\left(q^{2}\right) g^{\mu \nu}+\mathcal{O}\left(q^{\mu} q^{\nu}\right)$, and the correction terms in the lowest-order expressions are $s_{w}^{2}=g^{\prime 2} /\left(g^{2}+g^{\prime 2}\right)$ and $c_{w}^{2}=g^{2} /\left(g^{2}+g^{\prime 2}\right)$. 
Although the $\cal{S}, \cal{T},\cal{U}$ parameters are actually derived for the particles with a mass larger than the electroweak energy scale, we have checked at least for the mass of $W$-boson, which is enhanced by the ALP's loop diagrams, that the form of the oblique parameters is correct. Thus we adopt these expressions in our analysis.

\begin{table}[!t]
\setlength{\tabcolsep}{.5em}
\begin{tabular}{|c|c|c|c|}
\hline
  &  $C_{WW}/\Lambda$ &  $C_{BB}/\Lambda$ & $m_W$\\
\hline
PDG2021 &  -3.35  &  0.85 & $m_{W}=80.3819 \mathrm{GeV}$\\
\hline
PDG2021 &  3.34 &  -0.85 & $m_{W}=80.3819 \mathrm{GeV}$\\
\hline
Case A1 &  7.0 &  -0.4 & $m_{W}=80.4174 \mathrm{GeV}$\\
\hline
Case A2 &  -7.0 &  0.4 & $m_{W}=80.4174 \mathrm{GeV}$\\
\hline
Case B1 &  6.1 &  -0.5 & $m_{W}=80.4327  \mathrm{GeV}$\\
\hline
Case B2 &  -6.1 &  0.5 & $m_{W}=80.4327  \mathrm{GeV}$\\
\hline
\end{tabular}
\caption{Best-fitting results for ALP. The second and third columns are the best-fitting {\rm tan}($\alpha$) and $\xi$, respectively, for different values of the oblique parameters. The fourth column shows the corresponding values of $m_W$}.
\label{axion_bestfitting}
\end{table}

\begin{figure}
\centering
\includegraphics[width=0.95\linewidth]{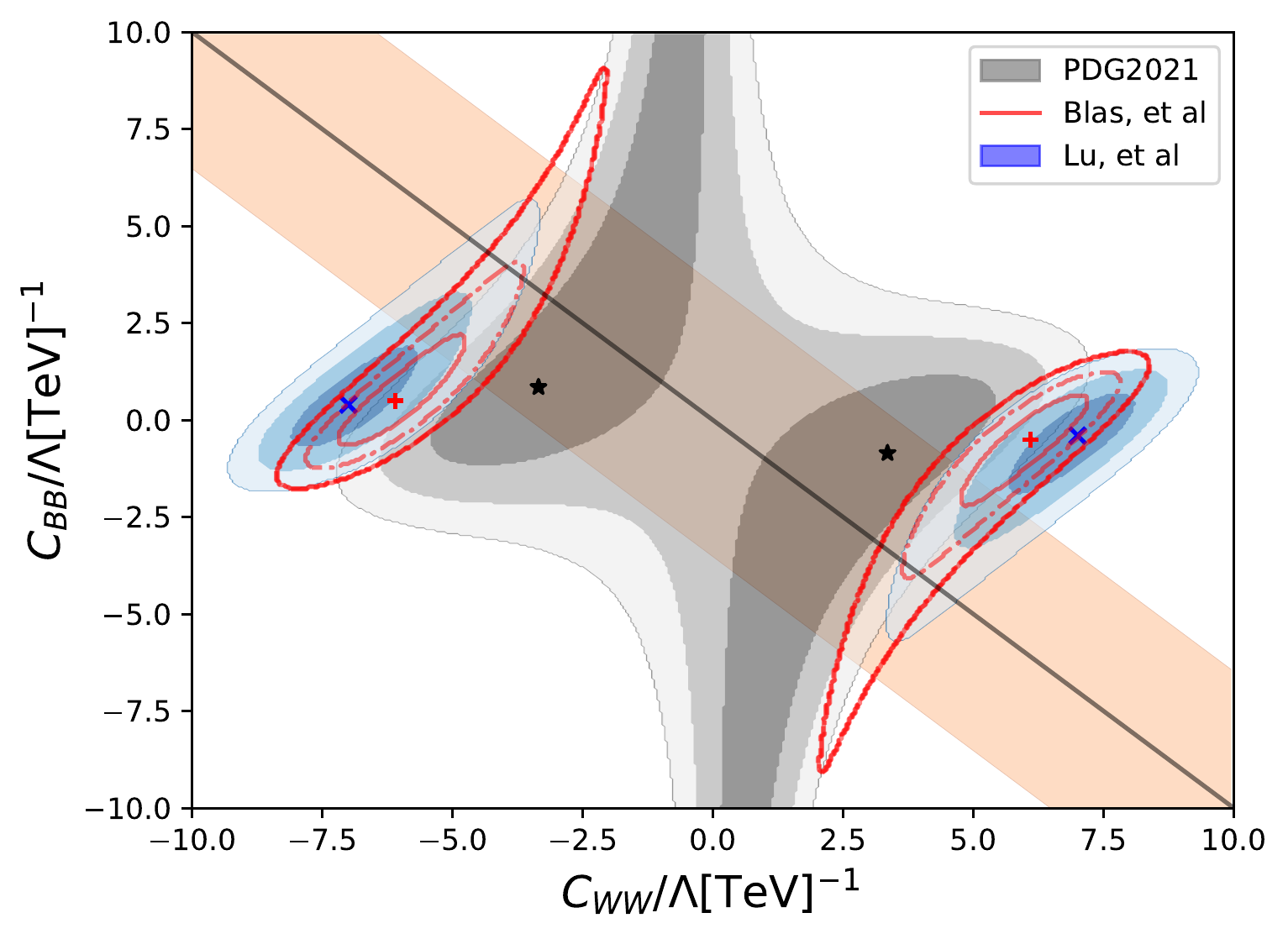}
\caption{Allowed parameter regions of ALP coefficients $C_{WW}-C_{BB}$ obtained from 2021 PDG results at 68\% (dark gray), 95\% (gray), and 99\% (light gray) CL.
As a comparison, we also plot the parameter space, adding the CDF $W$-boson mass, favored by global electroweak fit ${\cal S}$, ${\cal T}$, and ${\cal U}$ at 68\% , 95\%, and 99\% CL. The red contour lines and blue contours are for Case A and Case B, respectively. In addition, the orange region is allowed by the LEP collider at $M_a \sim 1$ GeV and the black line is set by the CAST observation~\cite{Bauer:2017ris}.
Our best-fitting results are marked with symbols.
} 
\label{axion_limit}
\end{figure}

With the $\chi^2$ defined in Eq.(\ref{chi2}), we obtain the allowed parameter spaces of Wilson coefficients in Fig.~\ref{axion_limit}. The best-fitting results are summarized in Table.~\ref{axion_bestfitting}. The results are shown in Fig.~\ref{axion_limit}, where the two red(blue) regions are favored by Case A (Case B) of CDF II. Our results with the previous 2021 PDG data are very similar to the results in \cite{Bauer:2017ris}, indicating that our method is sufficiently enough. Compared with the 2021 PDG results (i.e., the regions in grey), the viable regions are significantly narrowed down. 
Owing to such a paramount improvement, the current regions are stringently constrained by the CAST experiment \cite{Bauer:2017ris} for $m_a < {\cal O} {\rm (keV)}$ with $C_{\gamma\gamma} /\Lambda < {\cal O} {\rm (10^{-7}) TeV^{-1}}$ (i.e., the black line shown in Fig.~\ref{axion_limit}). Thus only a fine-tuning parameter space for $C_{\gamma\gamma} \sim 0$ is allowed. 
In other words, the interpretation of the $W$-boson mass excess with ALP is just marginally acceptable (i.e., it has been excluded at a significance level above $2\sigma$). Fortunately, for $m_a \sim {\rm GeV}$, we have a much weaker constraint of $C_{\gamma\gamma}/\Lambda < {\cal O}(1) {\rm TeV}^{-1}$ (shown in light orange).
The corresponding permitted region overlaps with the $\sim 2\sigma$ counters, suggesting that such massive ALP accounts for the correct $W$-boson mass without violating current observational constraints. Notably, the parameter regions with $m_a \sim \rm{GeV}$ and $C_{\gamma\gamma} /\Lambda \sim  \rm{TeV^{-1}}$ can also explain the muon magnetic moment anomaly \cite{Bauer:2017ris}. 
After analyzing the global electroweak fit result, we conclude that the new result of CDF has significantly narrowed down the allowed parameter regions of the ALP model. 

\subsection{Dark Photon}
The DP (or dark $Z^\prime$) is a gauge boson with an additional $U(1) ^\prime$ symmetry. 
It has been widely discussed for the experimental signatures by their kinetic mixing (with photon) or mass mixing (with $Z$-boson) \cite{Babu:1997st, Appelquist_2003, PhysRevD.44.2118, An:2012va, Deliyergiyev:2015oxa, Curciarello:2016jbz, Dobrescu:2004wz, Ackerman:2008kmp, Fabbrichesi:2020wbt, Caputo:2021eaa}.
In this study, we discuss the implication of the CDF II $W$-boson mass on this extending gauge structure of the SM broken at a weak scale.

The general renormalizable Lagrangian for the SM with an extra $U(1)^\prime$, neglecting the fermionic part, is given by \cite{Babu:1997st}:
\begin{equation}
\begin{aligned} 
\mathcal{L}_{Z^{\prime}}\supset &-\frac{1}{4} \hat{Z}_{\mu \nu}^{\prime} \hat{Z}^{\prime \mu \nu}+\frac{1}{2} M_{Z^{\prime}}^{2} \hat{Z}_{\mu}^{\prime} \hat{Z}^{\prime \mu}\\ 
&-\frac{\sin \alpha}{2} \hat{Z}_{\mu \nu}^{\prime} \hat{B}^{\mu \nu}+ \kappa\hat{Z}_{\mu}^{\prime} \hat{Z}^{\mu}, 
\end{aligned}
\end{equation}
where $\hat{B}_{\mu\nu}$ and $\hat{Z ^\prime}_{\mu\nu}$ denote the field strength tensors for $U(1)_{Y}$ and $U(1)^\prime$, respectively, $\hat{Z}$ is the $Z$-boson in the SM, and  $\hat{Z} ^\prime$ denotes the boson of the new $U(1)^\prime$, i.e., the DP (dark $Z ^\prime$). 
In addition, ${\rm sin}\alpha$ and $\kappa$ correspond to kinetic mixing with photon and mass mixing with $Z$-boson, respectively.
The physical eigenbasis can be obtained by diagonalizing the mass terms from $U(1)^{\prime}$ breaking and $SU(2)\times U(1)$ breaking. The nondiagonal mass terms give a mixing between $Z ^\prime$ and $Z$, thereby affecting the precision of electroweak measurements and the coupling with neutrinos \cite{Appelquist_2003,PhysRevD.44.2118,Fabbrichesi:2020wbt}.
In the end, one mass eigenstate is massless (the photon $A_{\mu}$), whereas the other two (denoted as $Z_{1}$ and $Z_{2}$) receive masses (we assume $m_{Z_1} < m_{Z_2}$, where $Z_1$ is the $Z$-boson in the SM, and $Z_2$ represents the dark $Z^\prime$). 
In terms of $(B_{\mu}, W_{\mu}^3, Z^{\prime}_{\mu})$, or alternatively $(\hat{A}_{\mu}, \hat{Z}_{\mu}, \hat{Z}^{\prime}_{\mu})$, the physical mass eigenstate shows:

\begin{equation}
\begin{aligned} A_{\mu} &=\hat{c}_{W} B_{\mu}+\hat{s}_{W} W_{\mu}^{3}=\hat{A}_{\mu}+\hat{c}_{W} \sin \chi \hat{Z}_{\mu}^{\prime} \\ Z_{1 \mu} &=\cos \xi\left(\hat{c}_{W} W_{\mu}^{3}-\hat{s}_{W} B_{\mu}\right)+\sin \xi Z_{\mu}^{\prime} \\ &=\cos \xi\left(\hat{Z}_{\mu}-\hat{s}_{W} \sin \chi \hat{Z}_{\mu}^{\prime}\right)+\sin \xi \cos \chi \hat{Z}_{\mu}^{\prime} \\ Z_{2 \mu} &=\cos \xi Z_{\mu}^{\prime}-\sin \xi\left(\hat{c}_{W} W_{\mu}^{3}-\hat{s}_{W} B_{\mu}\right) \\ &=\cos \xi \cos \chi \hat{Z}_{\mu}^{\prime}-\sin \xi\left(\hat{Z}_{\mu}-\hat{s}_{W} \sin \chi \hat{Z}_{\mu}^{\prime}\right) \end{aligned}
\end{equation}

\begin{figure}
\centering
\includegraphics[width=0.95\linewidth]{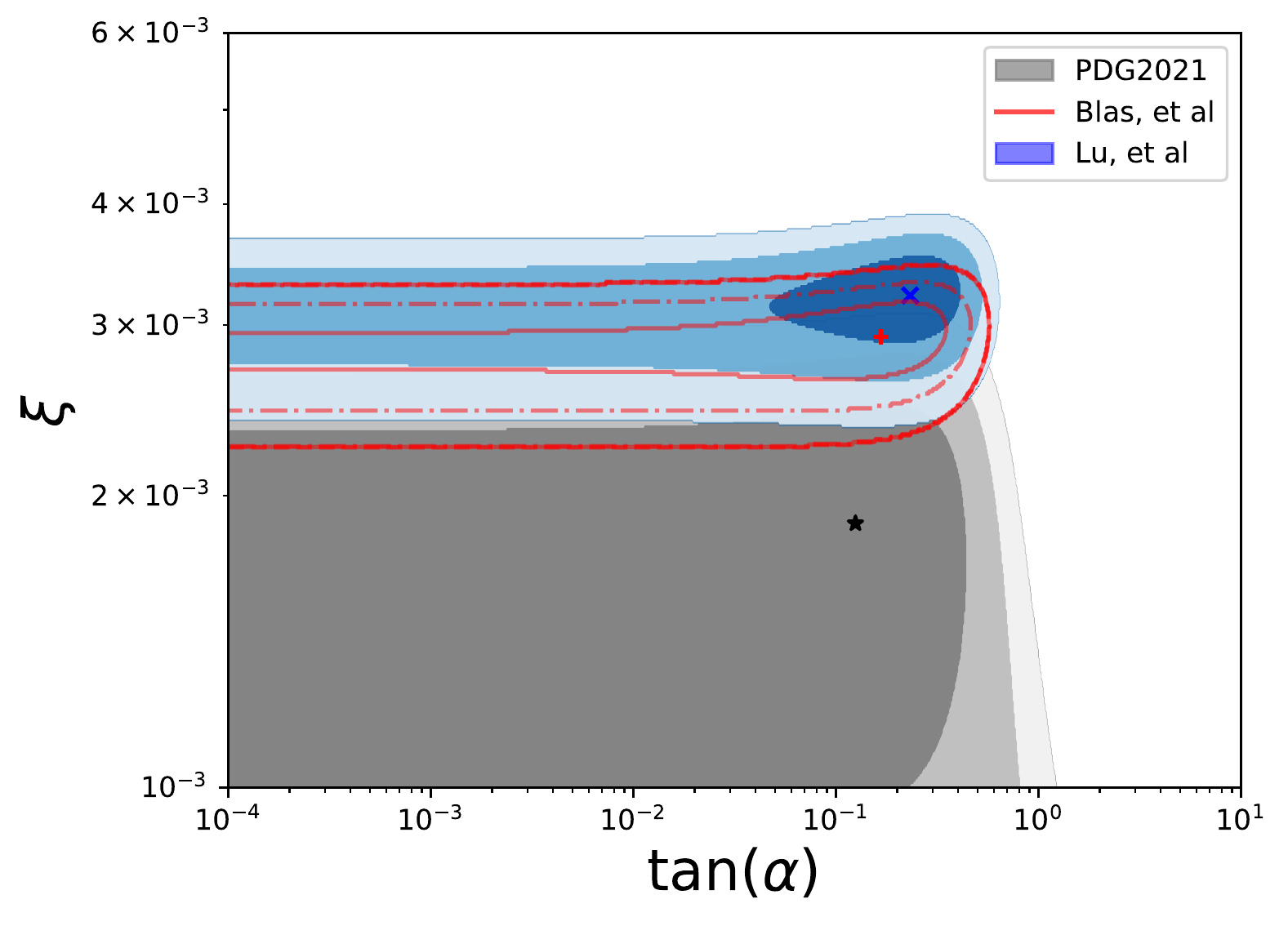}
\caption{Allowed parameter regions of dark photon obtained from 2021 PDG results at 68\% (dark gray), 95\% (gray) and 99\% (light gray) CL.
As a comparison, we also present the parameters space favored by global electroweak fit ${\cal S}$, ${\cal T}$ and ${\cal U}$ at 68\%, 95\%, and 99\% CL. Again, the red contour lines and blue contours are for Cases A and B, respectively. The symbols represent the best-fitting results.}
\label{darkphoton_limit}
\end{figure}

After defining
\begin{equation}
\tan 2 \xi=\frac{-2 \cos \alpha\left(\kappa+M_{Z}^{2} s_{W} \sin \alpha\right)}{M_{Z^{\prime}}^{2}-M_{Z}^{2} \cos ^{2} \alpha+M_{Z}^{2} s_{W}^{2} \sin ^{2} \alpha+2\kappa s_{W} \sin \alpha},
\end{equation} 
we can replace the mixing parameters ${\rm sin}\alpha$ and $\kappa$ with $\xi$ and ${\rm tan}\alpha$. Furthermore, the $Z-Z^{\prime}$ contribution to $\mathcal{S,~T,~U}$ (the leading order) could be expressed conveniently, which has been calculated in Ref.~\cite{Babu:1997st}. The DP enhances $m_W$ via kinetic mixing and mass mixing in our work.

\begin{table}[!t]
\setlength{\tabcolsep}{.5em}
\begin{tabular}{|c|c|c|c|}
\hline
  &  {\rm tan}($\alpha$) &  $\xi$ & $m_W$\\
\hline
PDG2021 &  $1.24\times 10^{-1}$ &  $1.87\times 10^{-3}$ & $m_{W}=80.3848 \mathrm{GeV}$\\
\hline
Case A &  $1.66\times 10^{-1}$ &  $2.92\times 10^{-3}$ & $m_{W}=80.4187  \mathrm{GeV}$\\
\hline
Case B &  $2.32\times 10^{-1}$ &  $3.22\times 10^{-3}$ & $m_{W}=80.4312 \mathrm{GeV}$\\
\hline
\end{tabular}
\caption{Best-fitting results for DP. The second and third columns are the best-fitting {\rm tan}($\alpha$) and $\xi$, respectively, for different values of the oblique parameters. The fourth column presents the corresponding $m_W$}.
\label{dark_photon_bestfitting}
\end{table}

In this work, we consider both the possible kinetic mixing and mass mixing terms. We choose two independent parameters, $\alpha$ and $\xi$, with a mass of $Z^\prime $ larger than 1 TeV. A stringent bound on the new vector boson has been set through the coupling with the SM leptons $g_{Z^\prime \bar{l}l}$.
The constraints are from $10^{-4}$ ($m_{Z^{\prime}} \sim \rm{GeV}$) to $0.1$ ($m_{Z^{\prime}} \sim \rm{TeV}$) \cite{ATLAS:2019erb,Hosseini:2022urq}. Thus, we only consider the case $m_{Z^{\prime}} \ge 1 \rm{TeV}$ to avoid the constraints from the collider in our work.  


With the above information and the new range ${\cal S,~T,~U}$ resulting in the global electroweak fit \cite{Lu:2022bgw}, we employ Eq.(\ref{chi2}) to calculate the favored regions and best-fitting results in the scenarios of Cases A and B (see the red contour lines and blue contours), and obtain $\xi \sim 3 \times 10^{-3}$ (see Fig.~\ref{darkphoton_limit} and Table.~\ref{dark_photon_bestfitting}).
For comparison, we also show the allowed regions with the PDG 2021 results. Clearly, similar to Fig.\ref{axion_limit}, the inclusion of the CDF II $W$-boson has significantly tightened the constraints, which is, in particular, the case of $\xi$, although the ranges of $\tan(\alpha)\leq 0.2$ are comparable with the previous ones. In the future, supposing the parameters of $\xi$ and $\tan(\alpha)$ have been independently constrained by other experiments or astrophysical data, the interpretation of the $W$-boson mass excess with DP will be further tested.

\subsection{Chameleon Dark Energy}
DE has been one of the most mysterious topics in modern cosmology since the discovery of the cosmic acceleration of the present universe \cite{Peebles:1987ek, SupernovaSearchTeam:1998fmf, Farrar:2003uw, Nojiri:2010wj,Bamba:2012cp,Joyce:2014kja, Cai:2015emx,Nojiri:2017ncd}. The prevailing dynamical model is to introduce a scalar field rolling along a flat potential so that negative pressure can be achieved. However, such a scalar field is beyond the SM and suffers from several conceptual issues, such as the equivalence principle \cite{Tino:2020nla}. One of the methods to release this tension is the so-called chameleon mechanism, which involves a scalar field with mass depending on the local matter density \cite{Khoury:2003rn, Khoury:2003aq, Brax:2004qh, Brax:2010xq, Burrage:2017qrf}.
Models involving these fields can give rise to acceleration naturally at late times in the universe, and remain consistent with local constraints. 
Such models have attracted considerable attention, as the requirement that the scalar field can vary with the local density of matter means that coupling to the SM states is compulsory. Therefore, in the early universe, chameleon fields are constrained by precision measurements, such as Big Bang Nucleosynthesis and the redshift of recombination. 
As the universe cools down, the background chameleon fields remain fixed in the minimum potential, with a slow drift in location. This induces a variation in the mass of any species of particles, that can couple with chameleon fields. 
These couplings imply that there may exist an interesting collider phenomenology. 

Precision electroweak measurements are screened from the indirect effect of DE, making such corrections effectively unobservable at present colliders, and limiting the DE discovery potential at CDF.
In this study, we shall choose to work with a theory of the broken phase of the electroweak force, in which the photon and the massive vector bosons could interact with chameleon scalar $\phi$ according to the action~\cite{Brax:2009aw}

\begin{widetext}
\begin{equation}
\begin{aligned} 
S=-\frac{1}{4} \int \mathrm{d}^{4} x\{& 2 \mathcal{F}(\phi)\left(\partial^{a} W^{+b}-\partial^{b} W^{+a}\right)\left(\partial_{a} W_{b}^{-}-\partial_{b} W_{a}^{-}\right)+4 m_{W}^{2} \mathcal{J}\left(\phi\right) W^{+a} W_{a}^{-} \\ &+\mathcal{F}(\phi)\left(\partial^{a} Z^{b}-\partial^{b} Z^{a}\right)\left(\partial_{a} Z_{b}-\partial_{b} Z_{a}\right)+2 m_{Z}^{2} \mathcal{J}\left(\phi\right) Z^{a} Z_{a} \\ &\left.+\mathcal{F}(\phi )\left(\partial^{a} A^{b}-\partial^{b} A^{a}\right)\left(\partial_{a} A_{b}-\partial_{b} A_{a}\right)\right\} ,
\end{aligned}
\end{equation}
\end{widetext}
where $W_a^{\pm}$ and $Z_a$ denote the gauge fields associated with the $W^{\pm}$ and $Z$, respectively, and $A_a$ denotes the gauge field associated with the photon. In addition, we introduce two functions $\mathcal{F}(\phi)$ and $\mathcal{J}(\phi)$, which describe how the chameleon scalar $\phi$ couples to the gauge boson kinetic and mass terms, namely,
\begin{equation}
\mathcal{F}(\phi)=\rm exp(\beta_{\gamma} \frac{\phi}{M_{\rm Pl}}),  \qquad
\mathcal{J}(\phi)=\rm exp(\beta_{m} \frac{\phi}{M_{\rm Pl}}),
\end{equation}
where $\beta_{\gamma}$ and $\beta_{m}$ denote the dimensionless chameleon couplings to photon and matter, respectively, and $M_{\rm Pl}$ denotes the reduced Planck mass.
The matter coupling $\beta_m$ is assumed to be universal to all species of matter, and the chameleon scalar is much smaller than the electroweak scale. The chameleon scalar $\phi$ can also enhance $m_W$ via loop diagrams, and its contributions are reflected in the self-energy correction.

\begin{figure}
\centering
\includegraphics[width=0.95\linewidth]{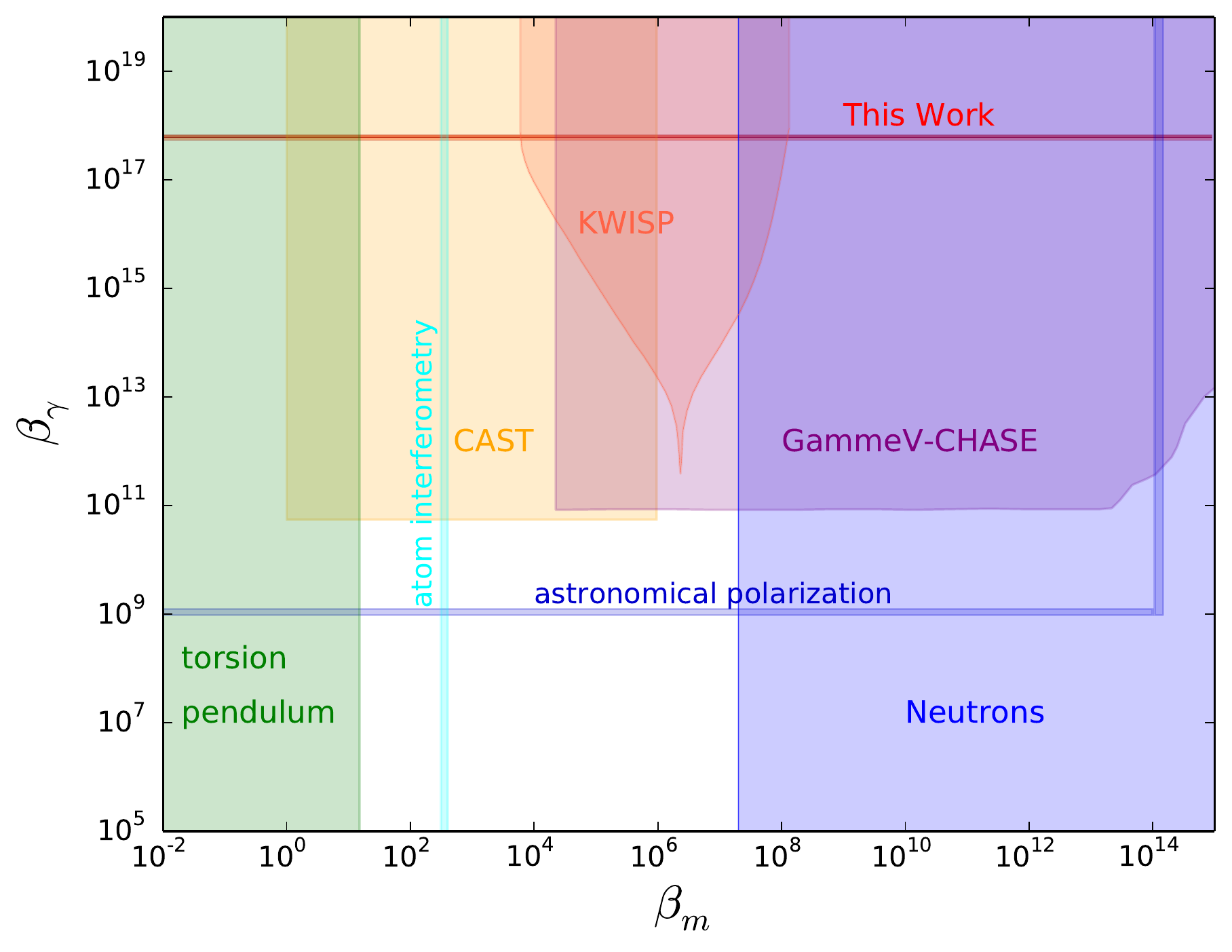}
\caption{Allowed parameter regions of chameleon dark energy obtained from a global electroweak fit  ${\cal S}$, ${\cal T}$, and ${\cal U}$ at 68\% (dark red), 95\% (red), and 99\% (light red) CL, the difference between Cases A and B can be ignored.
The constraints from other observations are also shown by different colors, including the torsion pendulum (green)~\cite{Upadhye:2012qu}, CAST in 2018 (orange)~\cite{CAST:2018bce}, CHASE (purple) ~\cite{Steffen:2010ze}, neutrons interferometry measurements (blue)~\cite{Lemmel:2015kwa} and KWISP (pink)~\cite{ArguedasCuendis:2019fxj}. 
In addition, the bounds of the atom-interferometry technique~\cite{Hamilton:2015zga, Jaffe:2016fsh} and the astronomical polarization~\cite{Burrage:2008ii} are represented with cyan line and dark blue lines, respectively. }
\label{chameleon_limit}
\end{figure}

In Ref.~\cite{Brax:2009aw} the forms of $\mathcal{S,~T,~U}$, were expressed in the terms of $\beta_m$ and $\beta_{\gamma}$ (please see Subsection 4.2 therein for the details). With the new ranges of $\mathcal{S,~T,~U}$ found in the recent global electroweak fit \cite{Lu:2022bgw}, now we can constrain the distribution of $\beta_m$ and $\beta_{\gamma}$ with Eq.\eqref{chi2}. The resulting favorable  parameter space is  presented in Fig.~\ref{chameleon_limit}, it is almost linear (see the red ribbon region).

Notably, the parameters of the chameleon DE can be independently constrained by various methods; in this study, we show some classic constraints. The torsion pendulum tests of the presence of new scalar forces yield a lower bound on the matter coupling $\beta_{m}$ \cite{Upadhye:2012qu}. The CAST experiment excludes large parameter regions by searching for solar chameleon at the soft X-ray energy range~\cite{CAST:2018bce}.
Neutron interferometry \cite{Lemmel:2015kwa} and atom-interferometry techniques~\cite{Hamilton:2015zga, Jaffe:2016fsh} yield upper limits, and provide a very strong bound. An additional astrophysical upper bound can be derived by analyzing the polarization of the light coming from astronomical objects \cite{Burrage:2008ii}. As shown in Fig.~\ref{chameleon_limit}, the possibility of attributing the $W$-boson mass excess to the presence of chameleon DE has already been convincingly ruled out by other experimental/astrophysical data.


\section{Summary and discussion}\label{conclusion}
The $W$-boson mass ($m_W= 80.4335\pm0.0094$GeV) measured by the CDF
collaboration is in excess of the SM prediction at a CL of 7$\sigma$, providing strong evidence for the presence of new physics beyond the SM. In the literature, various new particles or fields have been introduced to explain the DM or DE suggested by astrophysical observations. The presence of new particles or fields will modify the oblique parameters of ${\cal S}$, ${\cal T}$, and ${\cal U}$ and in principle can enhance the loop corrections of the $W$-boson mass. Thus, it is interesting to examine whether these widely-investigated scenarios are in tension with the new ${\cal S,~T}$, and ${\cal U}$ parameters found in the global electroweak fit including the CDF II $W$-boson mass measurement. For this purpose, we investigate three models involving either very light particles or scalar fields in this work, including the ALP, DP (dark $Z^{\prime}$), and chameleon DE models. Different from the relatively loose constraints on $C_{BB}$ and $C_{WW}$ with the 2021 PDG data, now the favored regions are separated and narrowly  distributed. 
The interpretation of the CDF II $m_W$ measurement with the ALP model is found to be just marginally consistent with the CAST constraint (see Fig.\ref{axion_limit}) except for $M_a \sim 1$ GeV.
Anyhow, the favored regions, though narrowly distributed, overlap with those inferred from the previous global electroweak fit (see Fig.~\ref{axion_limit}). Therefore, the CDF II $W$-mass measurement has played a significant role in narrowing down the parameter space of the ALP model, but the model is still viable. 
For the DP model, the latest global electroweak fit results yield a $\xi \sim 3\times 10^{-3}$ and $\tan(\alpha)\leq 0.2$ (see Fig.\ref{darkphoton_limit}) for $m_{Z^\prime} \gg O(EW)$. 
In the future, if independent constraints on these two parameters are yielded in other experimental/astrophysical data, this possibility can be further tested. 

For the chameleon DE model, we find an almost linear parameter region for the $m_W$ excess (see Fig.~\ref{chameleon_limit}). However, such a region has been convincingly excluded by the CAST, KWISP, and neutron interferometry experiments. 
In this study, we address the possible new light particle as the origin of the $W$-boson mass excess, and certainly there are many other interesting possibilities. 
For instance, it is argued that the inert two Higgs doublet model can naturally handle the CDF II W-boson mass without violating other experimental/astrophysical constraints, and the preferred DM mass is between 54 and 74 GeV \cite{Fan:2022dck}. In light of these facts, we conclude that the CDF II $W$-boson mass measurement can initiate a new window for the new physics, and significant progress is expected in the near future.

\section*{Acknowledgements} 
We thank Y. J. Wei, T. P. Tang, Z. Q. Shen, Z. Q. Xia, Y. L. S. Tsai, L. Wu and Q. Yuan for helpful discussion. 
This work is supported in part by the National Key R\&D Program of China (2021YFC2203100), by the NSFC (11921003, 11961131007, 11653002), by the Fundamental Research Funds for Central Universities, by the CSC Innovation Talent Funds, by the CAS project for young scientists in basic research (YSBR-006), and by the USTC Research Funds of the Double First-Class Initiative.

\bibliography{ref}

\end{document}